\documentclass[12pt,preprint]{aastex}
\shorttitle{Observations of M31-V1}
\shortauthors{Templeton et al.}
\received{2011 August 31}
\begin{document}
\title{Modern observations of Hubble's first-discovered Cepheid in M31}
\author{M. Templeton, A. Henden, W. Goff, S. Smith\altaffilmark{1}, R. Sabo, G. Walker,
R. Buchheim, G. Belcheva, T. Crawford, M. Cook, S. Dvorak, B. Harris}
\affil{American Association of Variable Star Observers, 49 Bay State Road, Cambridge, MA 02138; email: matthewt@aavso.org}
\altaffiltext{1}{also Robert Ferguson Observatory, Sugarloaf Ridge State Park, Kenwood, California}

\begin{abstract}
We present a modern ephemeris and modern light curve of the first-discovered
Cepheid variable in M31, Edwin Hubble's M31-V1.  Observers of the American
Association of Variable Star Observers undertook these observations during
the latter half of 2010.  The observations were in support of an outreach
program by the Space Telescope Science Institute's Hubble Heritage project,
but the resulting data are the first concentrated observations of M31-V1 
made in modern times.  AAVSO observers obtained 214 V-band, Rc-band, and
unfiltered observations from which a current ephemeris was derived.  The
ephemeris derived from these observations is 
$JD_{Max} = 2455430.5 (\pm 0.5) + 31.4 (\pm 0.1) E$.  The period derived from
the 2010 data are in agreement with the historic values of the period, but
the single season of data precludes a more precise determination of the period
or measurement of the period change
using these data alone.  However, using an ephemeris based upon the period
derived by Baade and Swope we are able to fit all of the observed data
acceptably well.
Continued observations in the modern era will be very valuable in linking
these modern data with data from the 1920s-30s and 1950s, and will enable us
to measure period change in this historic Cepheid.  In particular, we strongly
encourage intensive observations of this star around predicted times of
maximum to constrain the date of maximum to better than 0.5 days.
\end{abstract}

\keywords{Stars: variables: Cepheids -- Stars: individual: M31-V1 -- Galaxies: Local Group -- Galaxies: individual: M31 -- Astronomical databases: AAVSO -- {\it Facilities:} \facility{AAVSO}}

\section{Introduction}

The star M31-V1 (RA: 00:41:27.30, Dec: +41:10:10.4, J2000) was the first 
Cepheid discovered by Edwin Hubble in M31 \citep{Christianson1995}, thus 
making it an object of historical as well as astrophysical interest.  
The object was listed in \citet{Hubble1925} with a period of 31.41 days 
and a maximum magnitude of 18.2.  Hubble published his light curves in 
full in \citet{Hubble1929}, clearly demonstrating that this star and
several others like it in M31 exhibited Cepheid-like behavior.  After this
point, M31 was established as being a unique and distant Galaxy, independent
of the Milky Way.  Little or no 
work was done on these Cepheids until the work of W. Baade, published 
posthumously by H. Swope in 1965 \citep{BS1965}. Baade provided several 
additional years of data for M31-V1 along with many other variables.  
Baade's observations confirmed those of Hubble, and their analysis showed 
that the gross properties of M31-V1 were not significantly changed since the 
1920s.  Beyond these two data sets, little or no published time series 
photometry exists for M31-V1.

M31 has been extensively surveyed for Cepheids in both the spiral arms 
and disk \citep{Magnier1997,Kaluzny1998,Mochejska2000} and bulge 
\citep{Ansari2004}.  Surveys such as these give an increasingly complete 
snapshot of the population of Cepheids and other variables in M31, and 
consequently of the underlying stellar populations from which they 
arise.  What is not yet available for these variables is long temporal 
coverage with time-series data.  The majority of follow-ups of Hubble's 
and Baade \& Swope's work are photometric in nature, focused on 
calibration of the Leavitt Law \citep{Welch1986,FM1990} or on the 
general stellar population of M31 \citep{Massey2006} rather than on the 
individual stars as astrophysical objects for study.  This is 
understandable given the critical importance of the Leavitt Law in 
modern cosmology and the obvious difficulty of observing stars at the 
distance of M31, but it gives an incomplete picture of the potential 
value of M31 Cepheids to stellar astrophysics.  Measurements of period 
change in Cepheids have the potential to reveal more about the 
underlying stellar population and evolution of M31-V1 and other M31 
Cepheids \citep{TAB2006}, and where sufficient data exist to study 
individual Cepheids in detail, they should be utilized.

In 2010, observers with the American Association of Variable Star 
Observers (AAVSO) obtained just over 200 days of observations of M31-V1.  
These observations were in support of a public outreach project by the 
{\em Hubble Space Telescope} Hubble Heritage project, but the resulting 
data set is also a valuable new set of time series for M31-V1 in the 
current epoch, the first published time-series for this star in nearly 
sixty years.

In this paper, we present our analysis of these new time series 
observations of M31-V1, and place them in context of prior observations 
by Hubble and by Baade.  In Section 2 we describe the observers and 
observations, and in Section 3 we present the results of our analysis of 
these data.  In Section 4 we present the digitized light curves of the 
Hubble and Baade \& Swope data sets, and our attempt at phasing modern 
data with these older data sets.  In Section 5 we conclude with a 
discussion of our results and make a case for future 
observations.

\section{Observations}

In 2010 the {\em American Association of Variable Star Observers} 
(AAVSO) received a request for observations of this Cepheid by the 
Hubble Heritage project at STScI.  Observations of M31-V1 were 
tentatively planned with the {\em Hubble Space Telescope} to obtain 
time-series images intended for public outreach \citep{Noll2011}, but a 
modern ephemeris was unavailable.  Such observations would be 
non-trivial to plan at a ground-based observatory; an adequate ephemeris 
would require at least two cycles be observed with good signal to noise, 
and a sufficient number of observations per cycle to adequately detail 
the light curve.  For this star, that would require at least 60 days of 
coverage, with observations every 2-3 nights at least.  For this reason, 
the AAVSO observer community was asked to provide coverage of M31-V1 for 
several months prior to the planned HST visit in December 2010 and 
January 2011.

\begin{figure}
\figurenum{1}
\label{allaavsodat}
\epsscale{0.85}
\plotone{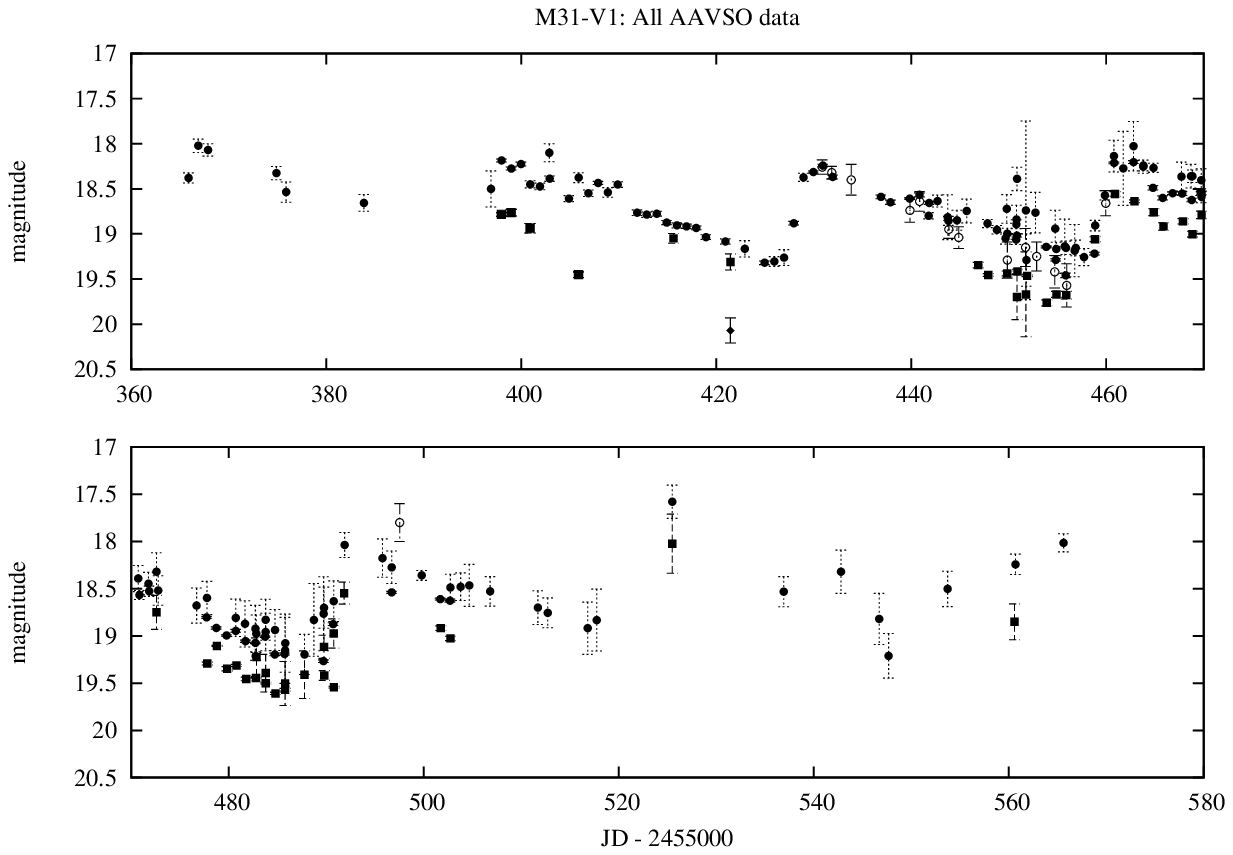}
\caption{The full AAVSO light curve of M31-V1 submitted by the eleven
participating observers.  Symbols: filled diamond, $B$ filter; filled square,
$V$ filter; open square, clear filter with $V$ zero-point; filled circle,
$R_{C}$ filter; open circle, clear filter with $R_{C}$ zerpoint.}
\end{figure}

M31-V1 has a magnitude around $V \simeq 19.4, (V-R) \simeq 0.6$
\citep{Massey2006}, 
which makes it a challenging but achievable target for amateur observers 
with larger telescopes and modern CCD cameras.  The AAVSO established a 
set of comparison stars specifically for this work that lie within 15 
arcminutes of M31-V1, using data from the USNO 1.0-meter and {\em 
Sonoita Research Observatory} 0.5-m telescopes.  After comparison stars 
were established and made available through our online chart plotter, 
the AAVSO issued a call for observations on July 16, 2010 
\citep{Waagen2010}.  Eleven different observers attempted observations 
of this target during the following six month period, and total of 214 
positive observations of M31-V1 were made between JD 2455365.9 and JD 
2455565.6 (2010 June 18 to 2011 January 4). The resulting light curve
of all AAVSO data is shown in Figure \ref{allaavsodat}.  An approximate
ephemeris was calculated using these data, and {\em HST} successfully 
imaged this region of M31 over several weeks, obtaining their desired data
set \citep{Soderblom2011}.

The eleven observers who contributed data to the HST campaign are given 
in Table \ref{observertable}.  The majority of the observations (152 of 
214 total) were obtained using Cousins R ($R_{C}$) filters, and our 
analysis is based upon these $R_{C}$-band observations.  Each observer
has a unique telescope and camera system; in general the observers were
able to reach a signal to noise of at least 3, but a few observers were
able to obtain very good signal to noise, with photometric accuracy
much better than 0.1 magnitudes per observation.  All of the observations
used in this paper are publicly available in the 
{\em AAVSO International Database}, and may be downloaded from the AAVSO's
website (http://www.aavso.org/data-download) using the name 
``{\bf M31{\textunderscore}V1}''.

\begin{figure}
\figurenum{2}
\label{figphasedr}
\epsscale{0.85}
\plotone{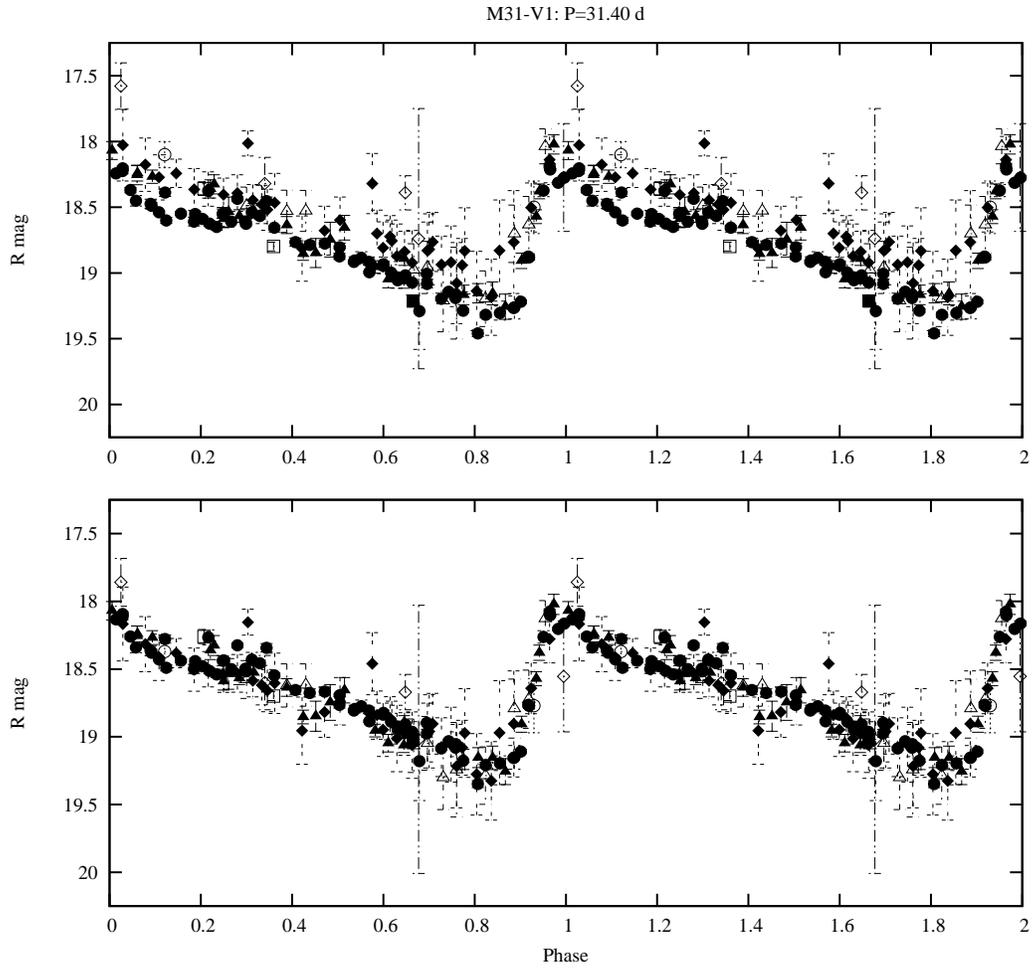}
\caption{$R$-band light curves of M31-V1 before (top) and after (bottom) 
zero-point offsets were applied. Symbols: filled squares, CMJA; open squares, CTX; filled circles, GFB; open circles, HBB; filled triangles, HQA (reference observer); open triangles, SRIC; filled diamonds, SSTB; open diamonds, WGR}
\end{figure}

The observations are not on a common standard system, but for the purpose of
time series analysis, this is not necessary.  Observers are using common
comparison 
stars and similar (if not identical) filters, which is sufficient.  For 
our analysis we simply require that the amplitude each observer observes 
is the same; all zero-point differences were removed prior to 
time-series analysis via an iterative procedure described below. To find 
the relative zero points of each observer, we first phased the data 
using a rough period of 31.41 days, obtained by Fourier transforming the 
raw data from a single observer (GFB, 65 $R_{C}$ observations).  The 
entire data set are then folded on that period, yielding a phased light 
curve of all observers.  We then divided the light curves into equal 
bins of 1/20 of a period, and performed an iterative adjustment of each 
observer's magnitude offset to minimize the sum of the variances of the 20 bins:

\begin{equation}
\sigma_{\Sigma} = \sum_{i=1}^{N_{b}} [ \sum_{j=1}^{N_{o}} {((m_{j,k}+\delta_{k}) - \bar{m})^{2}\over{N_{o}}}],
\end{equation}

\noindent
where $N_{b}$ is the number of bins, $N_{o}$ the number of observations per 
bin, $m_{j,k}$ the magnitude of the $j$-th observation by observer $k$ 
in each bin, and $\delta_{k}$ the magnitude offset of the $k$-th observer.
Each observer's offset $\delta_{k}$ was iteratively adjusted up or down by 
a maximum of 0.01 magnitudes per step, and the process was repeated 
until the sum of the variances of the bins reached a minimum.  These 
offsets were then applied to each observer's data, with observer HQA 
used as the reference magnitude; the offsets are given in Table 
\ref{offsettable}, and the raw and offset-fit phased light curves are 
shown in Figure \ref{figphasedr}.

\section{Analysis}

We performed a Fourier analysis on the resulting combined light 
curve to obtain the period, and phased the data to determine the time of 
maximum for the ephemeris.  For the Fourier analysis, we use an iterative
cleaning Fourier analysis code based upon the algorithm of \citet{RLD1987}
designed for irregularly-spaced observations, which is available from the
AAVSO website.
Fourier analysis of the light curve yields a peak at a frequency of 
$3.185 \times 10^{-2} d^{-1}$ (or $P=31.397 d$) with an $R_{C}$-band amplitude
of $0.35$ magnitudes.  The precision of the frequency measurement
is limited by the uncertainties in the observations and short temporal span
of the data.  The minimum $1-\sigma$ error on the frequency can be estimated 
using the relation for $\sigma(f)$ given by \citet{LB2005}, which depends
directly on the ratio of noise to amplitude, inversely on the span of the data
and the square root of the number of data points.
We assumed an average error per observation of 0.1 magnitudes, which yields
a frequency uncertainty measured from a single season of data (spanning 199.8
days) of about $1 \times 10^{-4} d^{-1}$, or $P = 31.4 \pm 0.1 d$.  This 
value matches what was derived from earlier data within uncertainties.
This is the {\it minimum} error that we expect on the period, and in reality
it could be significantly higher if we have underestimated the average 
uncertainties of the magnitude estimates, $\sigma$.

Phasing the data using this period yields a 
time-of-maximum of JD $2455430.5 \pm 0.5$ with the resulting linear 
ephemeris being

\begin{equation}
JD_{max} = 2455430.5(5) + 31.4(1) E.
\end{equation} 

We note that our method of establishing individual offsets may have
influenced the determination of the period slightly, because they were 
calculated assuming a phasing period for the light curve of 31.4 days.
However, we believe that this effect is minimal because the majority of 
observers obtained data at multiple phases of the light curve.  If the period
were greatly in error, it is likely that a single constant offset would not
decrease the scatter across all phases of the light curve as we have shown.
Since the bins are of order 1/20 of the period, we believe that the maximum
error in period would be about one bin width, 1.5 days.  Given that the period
was first estimated using a single observer's data, we believe that the error
in period is likely well below this, and is probably closer to the formal
error given above.

\subsection{Color curves}

\begin{figure}
\figurenum{3}
\label{fig_color}
\epsscale{0.85}
\plotone{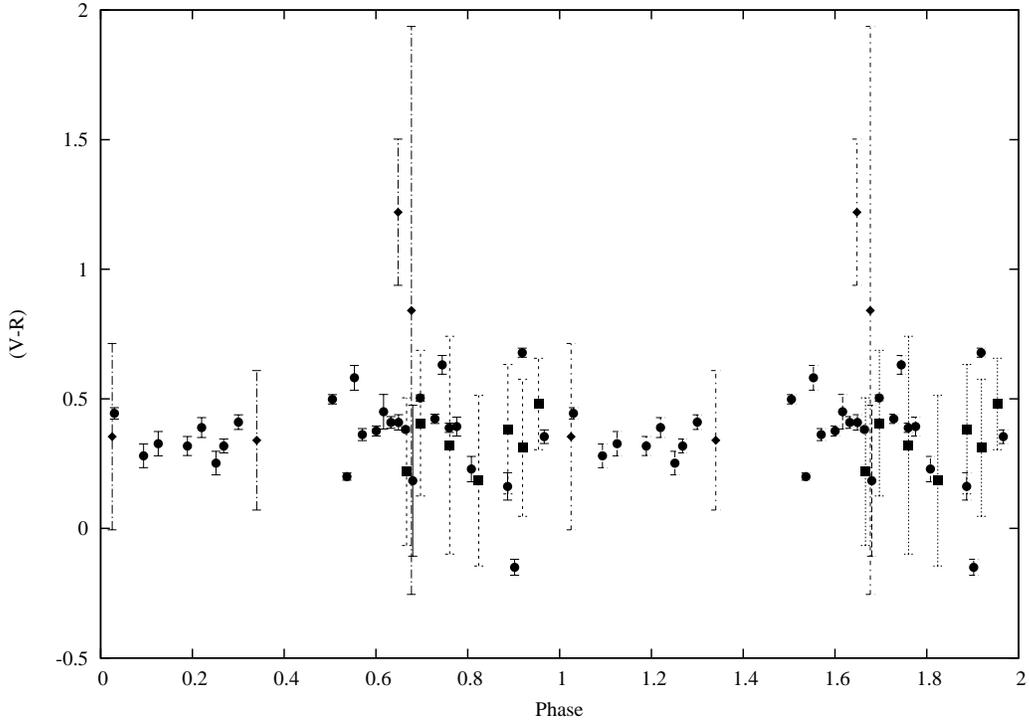}
\caption{The $(V-R)$ color curve of M31-V1 by three observers: GFB, filled
circles; SRIC, filled squares; WGR, filled diamonds.  The scatter in
the color measures is large, which precludes us from saying anything about
the color variations.}
\end{figure}

Both $R_{C}$ and $V$ band data were submitted by four different observers,
three of whom observed this star in both filters on at least one night.
We attempted to create color curves using data from these three observers
by phasing individual observers' nightly measures of $(V-R)$ with 
the ephemeris given above.  The resulting phased color curve is shown in
Figure \ref{fig_color}.  The phase curve is flat within the errors on
individual points, but the average value of $(V-R)$ is around +0.4,
which is somewhat bluer than is typical for Milky Way long period Cepheids
\citep{BT1995}.
However, we caution that while observers are using well-calibrated comparison
stars for photometry, the data for these observers was not fully calibrated
and transformed to a standard system.  The fact that there are zero-point 
$R_{C}$-band differences between observers suggests that there will be 
$V$-band differences as well.  Unfortunately there are no calibrated reference
observations in $V$ band as there are in $R_{C}$, which means that our
calculated average of $(V-R) = 0.4$ is not reliable.  However, we can
say that the variation in color over the cycle is smaller than we can measure
given the photometric errors.

\section{Historic Observations: Hubble \& Baade}

Both \citet{Hubble1929} and \citet{BS1965} published all of their photometry
in an easily extractable format, and we keypunched the observations from these
two papers into electronic, machine-readble tables.  These observations have
since been added to the {\em AAVSO International Database} and are freely
available
via the AAVSO website.  As part of this project, we wanted to compare how
the old data compare to the new by performing the same analyses on the archival
data as we did on our own photometry.  In this section, we present analyses of
the \citet{Hubble1929} and \citet{BS1965} data using the same procedure as
in Section 3.

\subsection{Hubble photometry}

\begin{figure}
\figurenum{4}
\label{hubbledat}
\epsscale{0.85}
\plotone{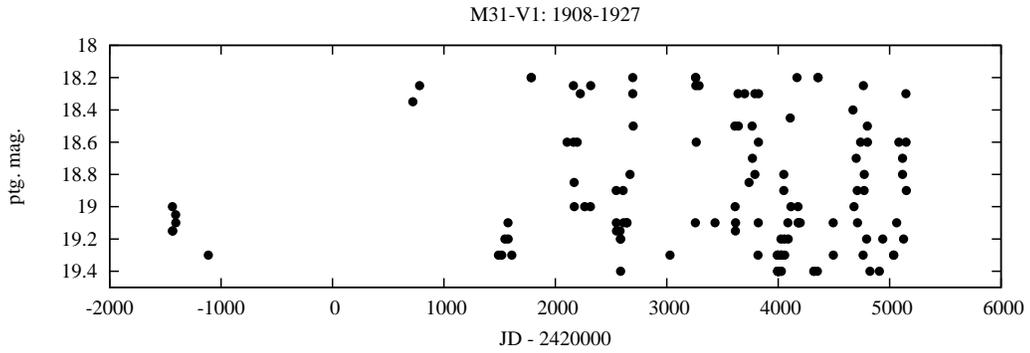}
\caption{Photographic light curve of M31-V1 from plates obtained with the
Mount Wilson 60- and 100-inch telescopes.  Measurements by E. Hubble
\citep{Hubble1929}.}
\end{figure}

Hubble's observations were described in \citet{Hubble1929}, and will be
briefly reviewed here.  The measurements were made from plates taken with
the Mount Wilson 60- and 100-inch telescopes by ten different observers
between JD 2418562.7 (1909 September 13) and 2425149.6 (1927 September 26).
We digitized 130 positive observations of M31-V1 by Hubble; we included all
observations regardless of Hubble's quality flag (``good", ``fair", or ``poor")
but we did not digitize observations where the star was below the faintest
camparison, or where it was not visible.  The resulting light curve is shown
in Figure \ref{hubbledat}.

\begin{figure}
\figurenum{5}
\label{hubblefold}
\epsscale{0.85}
\plotone{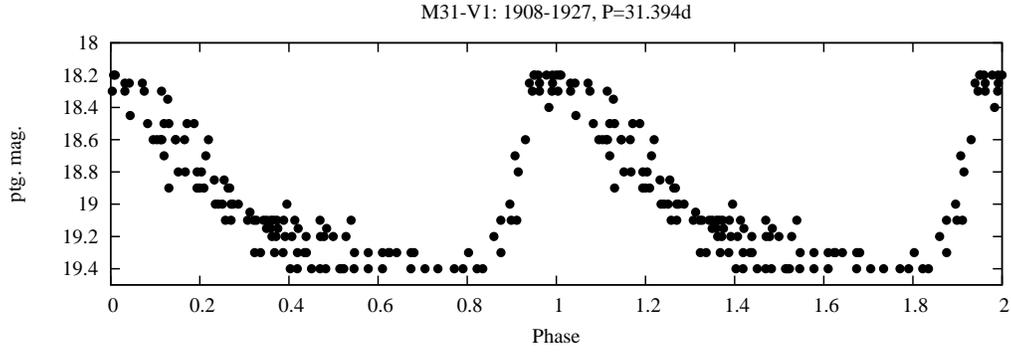}
\caption{Data shown in Figure \ref{hubbledat} folded with a period of 31.394
days.  We used $JD_{0} = 2422694.3$ to phase the data.}
\end{figure}

Fourier analysis of Hubble's data yields a period of $P = 31.394 d$, only 
slightly longer than Hubble's own measurement of $P = 31.390 d$.  The 
$1-\sigma$ error in frequency is $2.6 \times 10^{-6}$, corresponding to an
error in period of $\pm 0.003 d$; our measured period and that derived by
Hubble are identical within the uncertainties.  Hubble's data folded on
our period of $31.394 d$ is shown in Figure \ref{hubblefold}.  The full range
amplitude of
these data appears to be about the same as for the modern $R_{C}$ data,
around 1.5 magnitudes in the passband of the plates (assumed to be blue).
It is possible that Hubble's comparison stars weren't well-calibrated, 
particularly at the faint end, and it is therefore dangerous to make
assumptions about the variability by comparing the photographic light curve
to the modern $R_{C}$ curve.

\subsection{Baade photometry}

\begin{figure}
\figurenum{6}
\label{bsdat}
\epsscale{0.85}
\plotone{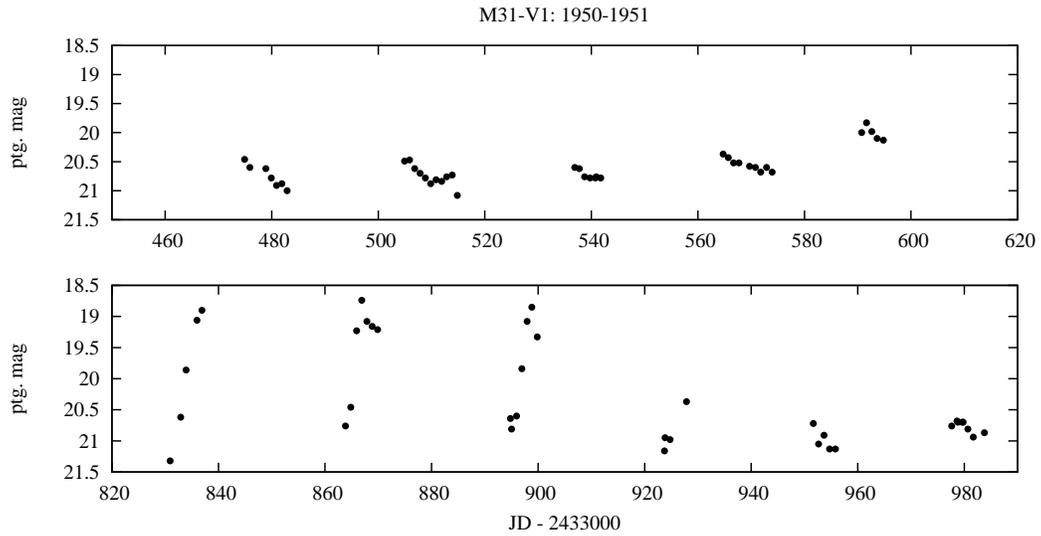}
\caption{Photographic light curve of M31-V1 from plates obtained with the
Palomar 200-inch telescope.  Plates by W. Baade, photometry by Baade and H.
Swope.}
\end{figure}

This star was observed by Walter Baade with the Palomar 200-inch in 1950 and
1951, although the data were not published until well after his death when
Henrietta Swope brought his work to light \citep{BS1965}.  Baade and Swope's
magnitudes are likely to be more accurate than Hubble's; \citet{BS1965} used 
photoelectric calibration star magnitudes (in SA 68) from W. Baum at Palomar and
\citet{Stebbins1950}, and Baade had much stricter control over the consistency
of plates used for imaging the variable and calibration fields.  The lightcurve
of Baade \& Swope's digitized data is shown in Figure \ref{bsdat}.

\begin{figure}
\figurenum{7}
\label{bsft}
\epsscale{0.85}
\plotone{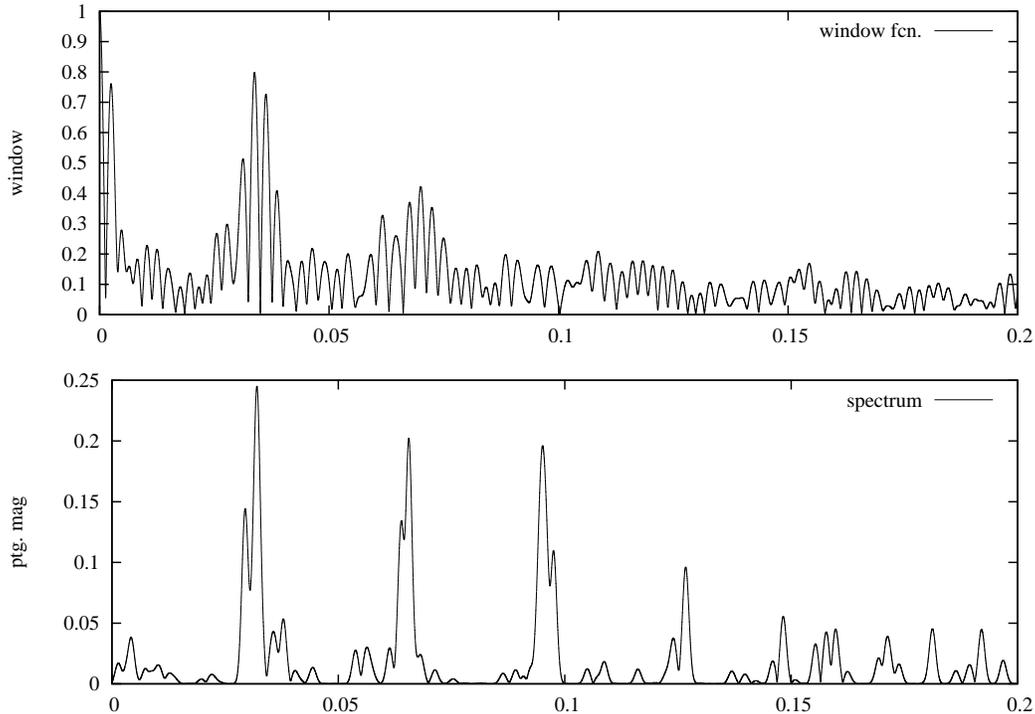}
\caption{Window function (top) and Fourier transform (bottom) of the 
\citet{BS1965} data set.  The window function shows clear signs of both
annual and monthly sidelobes.  The \citet{RLD1987} clearly did not adequately
clean the effects of the sidelobes, as can be seen in the double-peaked 
structure of the Fourier transform peaks.  For this reason, we consider our
period determination of these data to be unreliable.}
\end{figure}

Fourier analysis of these data yields a period of $31.226 d$ which is at
first glance much shorter than the Hubble data.  The Baade \& Swope data
span only two seasons (508.9 days), and are smaller in number than Hubble's
data but the difference in period is still substantial.  If we assume
the uncertainties in magnitude are around 0.05 mag, the uncertainty in
period is on the order of $\pm 0.03 d$ which means the periods are still
different by several sigma.  The likely explanation for this discrepancy
can be seen in the window function and Fourier transform of the Baade data,
which clearly show very strong annual and monthly sidelobes 
(Figure \ref{bsft}).  Based upon the dates of
observations, he was observing exclusively during dark time around new
moon.  The \citet{RLD1987} algorithm used for our Fourier calculations generally
works very well in most cases, but can still produce spurious results
when sidelobes strongly affect the transform.  The problem may also be
worsened given that the period of M31-V1 is on the order of a month which
can be seen in the similarity of the the star's observed phase from month
to month.

\begin{figure}
\figurenum{8}
\label{bsphase}
\epsscale{0.85}
\plotone{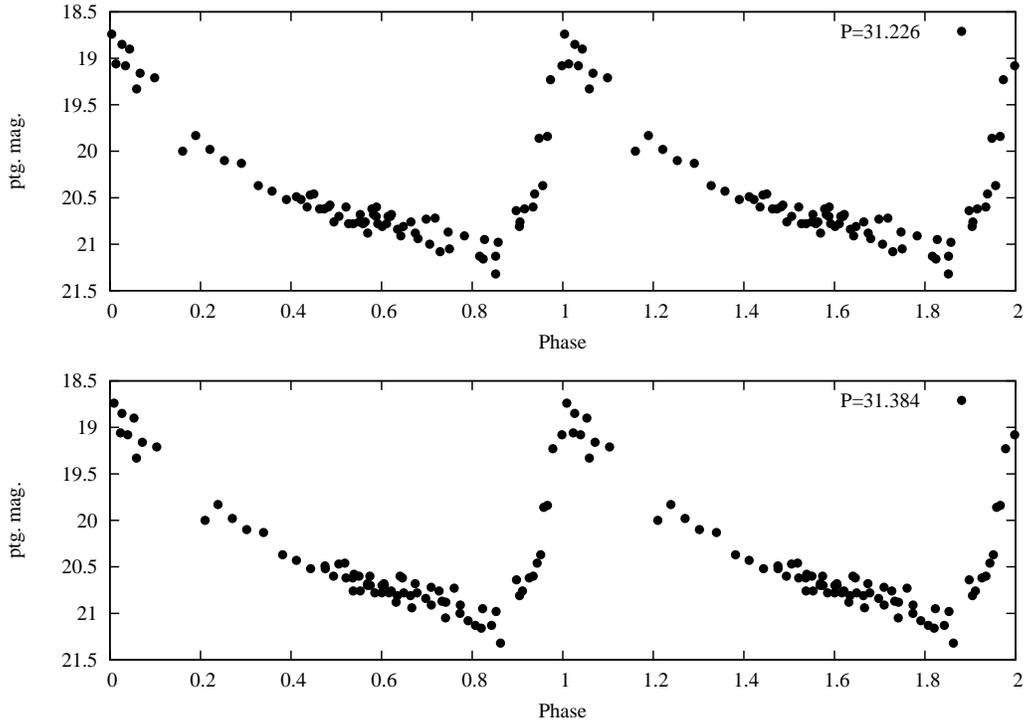}
\caption{Phased light curve of the Baade \& Swope data using two different
phasing periods.  Top: our period of 31.226 days; bottom: \citet{BS1965} period
of 31.384 days.  Both periods produce acceptably good phase curves, but we
believe our much shorter period is not correct on physical grounds that the
period should not have changed by that amount since the early 1900s.}
\end{figure}

\citet{BS1965} do not state how their period was calculated, but they
appear to have used the Hubble period as a starting point and simply adjusted
the period for a best fit.  They give a period $31.384 d$ for M31-V1, stating
that it gives a good fit of the earlier Hubble data as well their own.  We
show a phase curve of the Baade \& Swope data phased with our (likely
incorrect period) and that obtained by Swope in Figure \ref{bsphase}.  Both
periods produce acceptable phase curves, although the period from 
\citet{BS1965} appears slightly better.  We therefore accept the
\citet{BS1965} period and reject our own.  We used the $31.384 d$ period to
generate a test ephemeris with which we attempted to fit all measured historic
and modern data together, which we discuss below.

\subsection{Ephemeris for combined data}

\begin{figure}
\figurenum{9}
\label{tomfit}
\epsscale{0.85}
\plotone{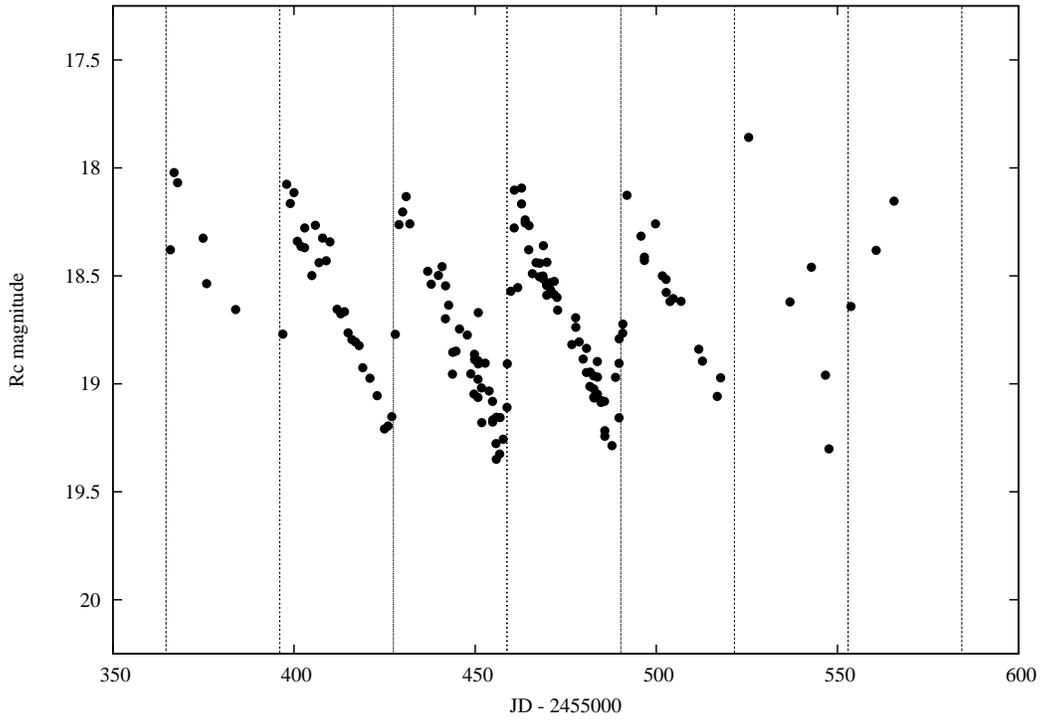}
\caption{Comparison of the 2010 $R_{C}$ band light curve with times of maximum
predicted using an ephemeris based on \citet{BS1965}.  The predicted times of
maximum are represented by the vertical lines.  The predicted times are very
close to the observed times, within 1-2 days, or about 6\% of the period.}
\end{figure}

As a final test, we generated an ephemeris for the entire data set from
1908 to 2010 using the 1950-1951 data as a central reference point.  We
used the period of $31.384 d$ from \citet{BS1965}, along with a 
well-measured time of maximum of JD 2433898.0 to generate predicted
times of maximum from 1908 to 2011.  In Figure \ref{tomfit} we show these
data overlaid onto the $R_{C}$ light curve.  This ephemeris produces a
reasonably good fit to the modern data, fitting the maxima to within 
about 5\% in phase.  Clearly an ephemeris derived from \citet{BS1965} can fit
the modern data acceptably well, suggesting that there has been no measurable
change in period during the entire course of observations.  However, this
cannot be guaranteed to be true, primarily because the three epochs of data
are very widely separated, and it is possible that different periods could
fit the data simply by adjusting both the period and the cycle number $E$ to
produce better fits.  Such a test will be conducted in a future paper planned
once another season of data is collected.

The primary reason that variables are observed over long periods of time or
in widely-spaced epochs is to look for changes in variations that provide
astrophysical insight into their evolution.  Some Cepheid variables are known
to change over time \citep{Szabados1983,TAB2006}, and these changes can yield
information about their position on the H-R diagram, evolutionary state,
and other physical properties.  The \citet{Hubble1929} and \citet{BS1965}
data sets are a possible starting point for studying period changes in
the M31 Cepheids.  The observations by AAVSO observers represent a
first attempt at detecting such changes serendipitously.  The single season
of data which we have obtained provides a preliminary constraint on period
change in M31-V1, and we expect that additional seasons of data during the
current epoch will provide more conclusive constraints.  In principle,
a similar modern study could be attempted on {\it any} Cepheid common to the
Hubble and Baade \& Swope samples, with the only possible limitation being
the available telescope time and the ability of telescopes of small aperture
to reach these faint magnitudes.  We suggest that any of the M31 Cepheids
with maxima brighter than $R_{C} = 19.0$ would be suitable targets for
future work.

\section{Discussion}

While the HST campaign was essentially a public outreach and education
initiative, the resulting data set from the AAVSO community represents a
new and valuable set of data for this star.  The quality of observations
suggests that the amateur community can do valuable work on the brighter
Cepheids in M31 or nearer galaxies in the local group.  The 2010 data set 
will hopefully be the first of several more years of follow-up observations.

The 2010 data set appears to be fully consistent with past observations of
M31-V1.  The evidence presented here -- while based on only a single
season of data -- is consistent with M31-V1 being physically
unchanged since its discovery by Hubble 80 years ago.  It is certainly
consistent with no significant period change since 1951, when
Baade last performed time-series photometry, however we caution that 
additional seasons of data will be required to confirm that definitively.
Subsequent years will enable us to more accurately define the period and
times of minimum, and we encourage the continued observation of M31-V1 by
the community, both amateur and professional.  We especially encourage
concentrated $R_{C}$-band observations around the times of maximum predicted
with the Baade \& Swope ephemeris.  With well-defined times of maximum we
can attempt to measure $(O-C)$ for this star, and should also be able to 
rule out the possibility that the period has changed significantly.

\acknowledgements
The corresponding author would like to extend his gratitude as always to the
observers who make this research possible; the AAVSO would not exist without
them.  We acknowledge with thanks the variable star observations from the 
{\em AAVSO International Database} contributed by observers worldwide and
used in this research.  Thanks also to the AAVSO Sequence Team for preparation
of a comparison star sequence.  We also thank the referee, Dr. John Percy, for
his very helpful comments that improved the focus of the paper.  This research
has made use of NASA's Astrophysics Data System Bibliographic Services.

{}

\begin{deluxetable}{ccccccccc}
\tablenum{1}
\tablecolumns{9}
\tablewidth{0pc}
\tablecaption{Observers contributing data to the campaign.}
\tablehead{\colhead{Obscode} & \colhead{Name} & \colhead{Location} & \colhead{Total} & \colhead{$R_{C}$} & \colhead{$V_{C}$} & \colhead{$CR$} & \colhead{$CV$} & \colhead{$B$}}
\startdata
GFB & Goff & California, US & 98 & 65 & 29 & ... & 4 & ...\\
SSTB & Smith & California, US & 38 & 38 & ... & ... & ... & ...\\
HQA & Henden & Arizona, US & 28 & 28 & ... & ... & ... & ...\\
SRIC & Sabo & Montana, US & 18 & 11 & 7 & ... & ... & ...\\
BHU & Buchheim & California, US & 13 & ... & ... & 13 & ... & ...\\
WGR & Walker & Massachusetts, US & 10 & 5 & 5 & ... & ... & ...\\
HBB & Harris & Florida, US & 3 & 2 & 1 & ... & ... & ...\\
BGU & Belcheva & Bulgaria & 2 & ... & 1 & ... & ... & 1\\
CTX & Crawford & Oregon, US & 2 & 2 & ... & ... & ... & ...\\
CMJA & Cook & Ontario, Canada & 1 & 1 & ... & ... & ... & ...\\
DKS & Dvorak & Florida, US & 1 & ... & ... & 1 & ... & ...\\
\enddata
\label{observertable}
\end{deluxetable}

\begin{deluxetable}{ccc}
\tablenum{2}
\tablecolumns{2}
\tablewidth{0pc}
\tablecaption{R-band offsets for the seven observers with more than one
R observation.  Offsets were computed by minimizing the variance
of the phased and binned light curve.  Offsets are in magnitudes relative
to the reference observer, HQA.}
\tablehead{\colhead{Obscode} & \colhead{mag. offset} & \colhead{N(obs)}}
\startdata
HQA & 0.0 & 28 \\
CTX & -0.10 & 2 \\
GFB & -0.11 & 65 \\
HBB & +0.27 & 2 \\
SRIC & +0.09 & 11 \\
SSTB & +0.14 & 38 \\
WGR & +0.29 & 5 \\
\enddata
\label{offsettable}
\end{deluxetable}

\end{document}